\documentclass[11pt]{article}       

\usepackage{graphicx}
\usepackage{mathptmx}      
\usepackage{bbm}
\usepackage{amsmath,amsfonts}
\usepackage{multirow}
\usepackage[margin=1in]{geometry}
\usepackage{setspace}

\newcommand{\ind}{\mathbbm{1}}

\begin{document}

\title{Design and analysis considerations for a sequentially randomized HIV prevention trial \thanks{Supported by United States National Institute of Health Grant 5U19HD089881 and U24HD089880}}


\author{David Benkeser\thanks{benkeser@emory.edu} \and
        Keith Horvath \and
        Cathy Reback \and
        Joshua Rusow \and
        Michael Hudgens
}




\maketitle

\begin{abstract}
TechStep is a randomized trial of a mobile health interventions targeted towards transgender adolescents. The interventions include a short message system, a mobile-optimized web application, and electronic counseling. The primary outcomes are  self-reported sexual risk behaviors and uptake of HIV preventing medication. In order that we may evaluate the efficacy of several different combinations of interventions, the trial has a sequentially randomized design. We use a causal framework to formalize the estimands of the primary and key secondary analyses of the TechStep trial data. Targeted minimum loss-based estimators of these quantities are described and studied in simulation. 
\end{abstract}

\section{Introduction}
\onehalfspacing
\label{intro}

Transgender individuals are estimated to make up between 0.39\% and 0.53\% of the US population \cite{crissman2017transgender,meerwijk2017transgender}, with higher percentages of high school students identifying as transgender \cite{johns2019transgender}. Transgender, gender non-conforming, and gender non-binary (referred to here collectively as trans) demonstrate elevated rates of risky sexual behaviors relative to their cisgender peers, leaving this population vulnerable to infection with human immunodeficiency virus (HIV) \cite{reisner2019situated}. A potential avenue for improvement in the health of trans youth is through mobile technology-based interventions. Such interventions circumvent the face-to-face interaction required for traditional health care delivery, and are therefore highly scalable. Moreover, technology-based interventions may be extremely low cost, confidential, and can cultivate communication amongst users, which may be important to trans youth who face high levels of stigma in healthcare and employment settings \cite{reisner2015legal,perez2018we}. It is therefore of interest to design mobile interventions tailored to trans youth, and to evaluate the efficacy of these interventions in preventing HIV infection. 

The TechStep study is an ongoing randomized controlled trial (RCT) of mHealth interventions in high-risk HIV-negative trans youths and young adults (15-24 years old). Three trans-specific mHealth interventions were developed for TechStep: a short messaging service (SMS or text), a mobile-optimized webapp, and an eCoaching intervention, which offers one-on-one risk behavior counseling administered through a mobile video conferencing service \cite{lentferink2017key,yousuf2019effectiveness}. The goal of the trial is to evaluate the efficacy of these interventions for reducing sexual risk behaviors (e.g., condomless anal or vaginal intercourse, engagement in sex work, sex while feeling the effects of alcohol or drugs) and increasing uptake of pre-exposure prophylaxis (PrEP).  

The availability of multiple intervention modalities complicates the design of an RCT to evaluate its efficacy. We must consider whether the mHealth interventions should be offered individually, all together, or in a sequential fashion. A priori eCoaching might be expected to be the most effective intervention of the three; however, from an implementation perspective, scalability may be limited due to its cost. These and other considerations led to a sequential multi-arm randomized trial (SMART) design. 

SMARTs randomize participants to an initial intervention and monitor participants after receipt of the intervention \cite{lavori2000design,murphy2005experimental,bembom2008analyzing,murphy2009screening,nahum2012experimental,chakraborty2013statistical}. Based on interim data, participants may become eligible for re-randomization to a new intervention. After receipt of the intervention, participants are monitored again. Based on these new data (and all previous data), participants may again become eligible for re-randomization to a new intervention. This process of monitoring and re-randomizing to different interventions continues until a pre-specified timepoint. The data generated from such a trial can be used to evaluate the efficacy of one or several \textit{treatment rules}, that is, rules for how participants should be assigned treatment at each time. 

SMART trials have great potential to advance HIV prevention science. Results of such trials can move researchers closer to personalized care for individuals, and enable evaluation of interventions that are feasible to implement. In this paper, we discuss the motivation for a SMART design in the TechStep trial, the statistical methods needed to provide a robust and efficient analysis of the trial, and a simulation study to assess the power of such the design to answer key questions of interest. We hope that this work will provide researchers with a better understanding for the motivation of SMART HIV prevention trials, as well as some ideas for how such trials can be planned and analyzed in practice. 

\section{Scientific questions and implications for design}
\label{sec:scientific_questions}

The TechStep study is a planned RCT utilizing mHealth interventions in an HIV-negative trans adolescent and young adult population (15 to 24 years of age). Three mHealth modalities for intervention were designed for use in the trial: a text message system, a mobile-optimized webapp, and an eCoaching intervention. The trial will have a total of nine months of follow-up with clinical assessments on participants taken at baseline and at three-, six-, and nine months after enrollment. In this section, we define the primary and key secondary questions for this study. 

\subsection{Primary question}
To identify a primary question of interest, we define our outcomes and interventions of interest. For the outcome, we focus on the cumulative number of risky sexual encounters, defined as acts of condomless anal intercourse, engagement in sex work, or sex while feeling the effects of alcohol or drugs. We will consider an intervention effective if it reduces the \emph{average} number of risky encounters over the duration of the trial or increases PrEP uptake. 

Ideally, a primary analysis should assess an intervention that is feasible to implement in a broader clinical setting \cite{brown2017overview}. We reasoned that many clinics could offer either a the text- or an webapp-based intervention, but not both. Moreover, such low intensity interventions may be sufficiently efficacious for many participants in reducing risky sexual behaviors. However, there may be individuals for whom the low intensity interventions would be insufficient to change risk behaviors. These individuals may derive greater benefit from the more intensive and personalized eCoaching intervention. For these reasons, TechStep considered three interventions: (i) control intervention consisting of a static website with trans-specific HIV and risk reduction information and local and national trans resources, (ii) text + eCoaching, where participants receive text messaging for ninety days and those who do not improve their risk behaviors receive three months of eCoaching in addition to text messaging, and (iii) webapp + eCoaching, where participants receive access to a webapp for ninety days and those who do not improve receive eCoaching for three months in addition to the webapp. The primary goal of the trial is to compare average number of risky encounters for participants who receive intervention (ii) vs. (i) and who receive intervention (iii) vs. (i). 

\subsection{Key secondary questions}

Some health clinics may be unable to staff trained eCoaching counselors, and thus interventions (ii) and (iii) may be infeasible to implement in some clinical settings. However, clinics could likely support either a text message system or a webapp, since these interventions are extremely low cost. Therefore, we wanted to evaluate the efficacy of two additional interventions: (iia) text messaging only, where participants receive text messaging for six months and (iiia) webapp only, where participants receive access to a mobile-optimized webapp for six months. A key secondary goal of the trial is to compare the average number of risky encounters for participants who receive intervention (iia) vs. (i) and who receive intervention (iiia) vs. (i). 

While an eCoaching intervention is the most intensive and costly mHealth intervention we considered, it is also hypothesized a-priori to be the most effective. Thus, it is also relevant to assess the marginal benefit of offering eCoaching to participants whose behavior does not improve under other mHealth interventions. To this end, another key secondary goal of the trial is to compare the average number of risk encounters for participants who receive intervention (ii) vs. (iia) and (iii) vs. (iiia).

\section{Causal Estimands}

In this section, we introduce general notation to describe data collected in the TechStep trial, translate the primary and secondary questions into formal causal estimands and establish their identifiability based on the observed trial data. We use $L_0 = (Y_0, W_0)$ to denote a vector of baseline participant information collected at the initial clinic visit. We use $Y_0$ to denote sexual risk behaviors and $W_0$ to denote other (e.g., demographic) information collected. While in the actual TechStep trial, $Y_0$ will consist of multiple measures of risk behaviors, for simplicity, here we consider that $Y_0$ is scalar-valued. We use $A_0$ to denote the intervention assigned to participants at enrollment, with $A_0 = 0$ denoting that standard of care was assigned, $A_0 = 1$ denoting that a first-line text intervention was assigned, and $A_0 = 2$ denoting that a first-line webapp intervention was assigned. We use $C_1$ to denote whether a participant attends the three-month clinic visit, so that $C_1 = 0$ denotes that the participant did not attend the visit, and $C_1 = 1$ denotes that the participant did attend the visit. For participants who attend the three-month visit, we measure additional information on sexual risk behaviors over the last three months. In particular, $L_1 = (Y_1, W_1)$ denotes a vector of participant information collected at the three-month visit, where, as above, we use $Y_1$ to denote sexual risk behaviors over the first three months of the study, and $W_1$ denotes other information collected on participants. We use $A_1$ to denote the intervention assigned to participants at the three-month visit. For simplicity, here we consider that participants with $Y_1 \ge Y_0$ and $Y_1 \ne 0$, i.e., those who have at least some risky behaviors and whose behaviors do not improve under the first-line intervention, are re-randomized to either maintain their current intervention or to ``step-up'' to eCoaching. We set $A_1 = 3$ if a participant is chosen to step-up and set $A_1 = A_0$ otherwise. Participants who have no risky behaviors or whose behaviors have improved under first-line intervention maintain their current intervention, $A_1 = A_0$. 

We use $C_2$ to denote whether a participant attends the six month clinic visit and $L_2 = (Y_2, W_2)$ to denote information collected at this visit. Similarly, we define $C_3$ and $Y_3$ for the nine month clinic visit. We denote by $Y = Y_1 + Y_2 + Y_3$ the cumulative outcome over the duration of the study. Thus, the time-ordered data for a typical study participant can be represented as a random variable $O = (L_0, A_0, C_1, L_1, Y_1, A_1, C_2, L_2, Y_2, C_3, Y_3)$. We denote by $\mathbb{P}$ the distribution of the observed data unit $O$ and assume a nonparametric model for $\mathbb{P}$. 

To define our causal question of interest, we use a nonparametric structural equation model (NPSEM). This model assumes that each observed variable is a function of its parent variables and an exogenous error term. We adopt a model that assumes the parents of a given observed variable are all variables that temporally precede it. For example, $L_0$ has no parent variables, the parent of $A_0$ is $L_0$, the parent of $C_1$ is $A_0$ and $L_0$, and so on. We define interventions on this NPSEM that correspond to interventions (i), (ii), (iia), (iii), and (iiia) above. In addition to intervening on the mHealth variables ($A_0$ and $A_1$), these interventions enforce that participants attend each clinic visit. We define a function $d$ that maps $A_0,Y_0,Y_1$ to $\{0,1,2,3\}$, and defines our rules for stepping people up to the eCoaching intervention, \[
  d(a_0, y_0, y_1) = \left\{  \begin{array}{ll} 0 & \mbox{if $a_0 = 0$} \\
                                                1 & \mbox{if $a_0 = 1$ and ($y_1 < y_0$ or $y_1 = 0$)} \\
                                                2 & \mbox{if $a_0 = 2$ and ($y_1 < y_0$ or $y_1 = 0$)} \\
                                                3 & \mbox{if $a_0 \ne 0$ and $y_1 \ge y_0$ and $y_1 \ne 0$}
                              \end{array}  \right.
\]

The interventions of interest are shown in Table \ref{npsem_interventions}. We denote by $Y^{\text{(i)}}$, $Y^{\text{(ii)}}$, $Y^{\text{(iia)}}$, $Y^{\text{(iii)}}$, $Y^{\text{(iiia)}}$ the counterfactual outcomes generated by interventions (i), (ii), (iia), (iii), and (iiia), respectively. For example, $Y_j^{\text{(i)}}$ represents the total number of reported risky sexual behaviors over the duration of the trial, if, possibly counter to fact, individual $j$ received standard of care throughout the trial. We use $Y^{\text{($\cdot$)}}$ to generically denote one of these counterfactual outcomes. The distribution of the counterfactual variable $Y^{\text{($\cdot$)}}$ is denoted by $\mathbb{P}^{\text{($\cdot$)}}$. We are interested in $\psi^{\text{($\cdot$)}} = \mathbb{E}(Y^{\text{($\cdot$)}}) = \int y d\mathbb{P}^{\text{($\cdot$)}}(y)$ for each intervention. Our primary and secondary questions can be answered using statistical inference about parameters of the counterfactual outcome distributions. In particular, we quantify intervention effects in terms of additive differences in the average number of mean outcome over the duration of the trial. The causal estimands of interest pertaining to each of our primary and secondary questions are shown in Table \ref{causal_estimands}. 

\begin{table}
\begin{tabular}{lll}
\hline
Label & NPSEM intervention & Description \\
\hline
\multirow{2}{*}{(i)}  & \multirow{2}{*}{$A_0 = 0, C_1 = 1, A_1 = 0, C_2 = 1, C_3 = 1$} & standard of care for 6 months \\
     &                                               & and attend each visit \\
\hline
     & $A_0 = 1, C_1 = 1$                            & text messages for 3 months, step up  \\
(ii) & $A_1 = d(1, Y_0, Y_1)$ &  to eCoaching if no improvement at 6 \\
     & $C_2 = 1, C_3 = 1$                                        &  months and attend each visit \\
\hline
\multirow{2}{*}{(iia)}     & \multirow{2}{*}{$A_0 = 1, C_1 = 1, A_1 = 1$, $C_2 = 1, C_3 = 1$} & text messages for 6 months \\
                           &                 & and attend each clinic visit \\
\hline
     & $A_0 = 2, C_1 = 1$                            & webapp for 3 months, step up  \\
(iii) & $A_1 = d(2, Y_0, Y_1)$ &  to eCoaching if no improvement at 6 \\
     & $C_2 = 1, C_3 = 1$                                        &  months and attend each visit \\
\hline
\multirow{2}{*}{(iiia)}     & \multirow{2}{*}{$A_0 = 2, C_1 = 1, A_1 = 2$, $C_2 = 1, C_3 = 1$} & webapp for 6 months \\
                           &                 & and attend each clinic visit \\
\hline
\end{tabular}
\caption{Hypothetical interventions on nonparametric structural equation model.}
\label{npsem_interventions}
\end{table}

\begin{table}
\begin{tabular}{lll}
\hline
Question & Causal estimand & Description \\
\hline
\multirow{4}{*}{Primary} & \multirow{2}{*}{$\mathbb{E}(Y^{\text{(ii)}}) - \mathbb{E}(Y^{\text{(i)}})$} & difference in mean outcome under text \\
                         & &  \ \ + step intervention versus standard of care \\ \cline{2-3}
& \multirow{2}{*}{$\mathbb{E}(Y^{\text{(iii)}}) - \mathbb{E}(Y^{\text{(i)}})$} & difference in mean outcome under webapp \\
& & \ \ + step intervention versus standard of care \\ 
\hline
\multirow{8}{*}{Secondary} & \multirow{2}{*}{$\mathbb{E}(Y^{\text{(iia)}}) - \mathbb{E}(Y^{\text{(i)}})$} & difference in mean outcome under text-only \\
                         & &  \ \ versus standard of care \\ \cline{2-3}
  & \multirow{2}{*}{$\mathbb{E}(Y^{\text{(iiia)}}) - \mathbb{E}(Y^{\text{(i)}})$} & difference in mean outcome under webapp-only \\
                         & &  \ \ versus standard of care \\ \cline{2-3}
  & \multirow{2}{*}{$\mathbb{E}(Y^{\text{(ii)}}) - \mathbb{E}(Y^{\text{(iia)}})$} & difference in mean outcome under text \\
                         & &  \ \ + step versus text only \\ \cline{2-3}                        
  & \multirow{2}{*}{$\mathbb{E}(Y^{\text{(iii)}}) - \mathbb{E}(Y^{\text{(iiia)}})$} & difference in mean outcome under webapp \\
& & \ \ + step intervention versus webapp only \\ 
\hline
\end{tabular}
\caption{Causal estimands used to answer primary and secondary questions of interest.}
\label{causal_estimands}
\end{table}

In this section, we discuss identification of $\mathbb{E}(Y^{\text{(ii)}})$ using $\mathbb{P}$, the distribution of the observed data. Nearly identical arguments can be used to establish identifiability of the other causal estimands. We use the longitudinal G-formula to establish identifiability; see \cite{robins1999ltmle,bangrobins:2005:biometrics,vdlgruber:2012:ijb} for more on the assumptions under which this identification result holds. We draw particular attention to the assumption of sequential randomization of intervention and missingness, also referred to as the assumption of no unmeasured (time-varying) confounding. The SMART design of the trial ensures that intervention rules are randomly assigned and thus that observed intervention values are independent of counterfactual outcomes. However, as with any clinical trial, we expect that participants will miss clinic visits. Thus, in order that the sequential randomization assumption is satisfied, we require that amongst those participants who receive intervention (ii), $C_t$ is independent of $Y^{\text{(ii)}}$ given $\bar{L}_{t-1}$, the covariate information collected up to time $t-1$. That is, the covariate history $\bar{L}_{t-1}$ should include all individual characteristics that are related both to participants' propensity for risky sexual behaviors and for attending future clinic visits. We will adopt this bar notation throughout to denote a collection past measurements of particular variables. For example, $\bar{C}_{t} = (C_1, \dots, C_t)$ for $t = 1, 2, 3$. We will sometimes write $\bar{1}$ to denote a unit vector, the length of which will be apparent from the context. For example, $\bar{C}_3 = \bar{1}$ denotes the event that $C_1 = 1, C_2 = 1, C_3 = 1$, while $\bar{C}_2 = \bar{1}$ denotes the event that $C_1 = 1, C_2 = 1$, and so on.  

If the assumptions described above are satisfied, then $\mathbb{E}(Y^{\text{(ii)}})$ can be written as a functional of $\mathbb{P}$, for which we require the following definitions. We define $Q_3(y,\bar{\ell}_2)$ as the conditional CDF of $Y$ given $A_0 = 1, A_1 = h(1, y_0, y_1), \bar{C}_3 = \bar{1}$, and $\bar{L}_2 = \bar{\ell}_2$ evaluated at $y$. In words, $Q_3(y,\bar{\ell}_2)$ describes the probability of having fewer than or equal to $y$ risky sexual encounters over the duration of the trial, amongst the participants in intervention arm $\text{(ii)}$ that attended each of the three follow-up visits and had observed covariate history $\bar{\ell_2}$. Similarly, we define $Q_2(\bar{\ell}_2)$ as the conditional CDF of $L_2$ given $A_0 = 1, A_1 = h(1, y_0, y_1), \bar{C}_2 = \bar{1}, \bar{L}_1 = \bar{\ell}_1$ evaluated at $\ell_2$, and we define $Q_1(\bar{\ell}_1)$ as the conditional CDF of $L_1$ given $A_0 = 1, C_1 = 1$, and $L_0 = \ell_0$. Finally, we define $Q_0(\ell_0)$ as the conditional CDF of $L_0$ evaluated at $\ell_0$. The G-formula for the causal estimand of interest is \begin{equation*} \label{Qbar0}
\mathbb{E}(Y^{\text{(ii)}}) = \int \bar{Q}_1(\ell_0) dQ_0(\ell_0) \ , 
\end{equation*}
where $\bar{Q}_1$ is recursively defined as follows, \begin{align}
\bar{Q}_1(\ell_0) &= \int \bar{Q}_2(\bar{\ell}_1) \hspace{0.03in} dQ_1(\bar{\ell}_1) \label{Qbar1} \\
\bar{Q}_2(\bar{\ell}_1) &= \int \bar{Q}_3(\bar{\ell}_2) \hspace{0.03in} dQ_2(\bar{\ell}_2) \label{Qbar2} \\
\bar{Q}_3(\bar{\ell_2}) &= \int y \hspace{0.03in} dQ_3(y, \bar{\ell}_2) \ . \label{Qbar3}
\end{align}

\section{Statistical Methodology}

Well-controlled randomized trials generally allow researchers to utilize covariate-unadjusted statistical methods to draw inference on causal effects, since, by virtue of the randomized design, counterfactual outcomes are marginally independent of intervention assignment. However, the same is not true in SMART trials, as sequential randomization probabilities depend on past covariates. Thus, these covariates must be accounted for by the estimator in order to draw valid inference. A more challenging issue is that typically at least some follow-up data are missing in randomized trials, for example, due to participants moving out of the study region.  Simple strategies, such as complete-case analysis, run the risk of incurring bias due to informative missingness. To draw more robust inference, we can utilize estimators that account for {\it all} measured covariates, not just those that are used to determine re-randomization probabilities. If the measured covariates are rich enough to satisfy the sequential randomization assumption on missingness, then we should be able to draw valid inference about intervention effects, in spite of the informative nature of the missingness. 

We will analyze the TechStep trial using targeted minimum loss-based estimation (TMLE) \cite{van2006targeted,van2011targeted,van2018targetedbook}. TMLE is a general framework for generating plug-in estimators that solve a set of user-selected equations \cite{van2018targeted}. In the present problem, we sequentially generate estimates of the parameters of the G-formula (\ref{Qbar1})-(\ref{Qbar3}) that solve the efficient influence function estimating equation. We refer interested readers to \cite{petersen2014targeted,schnitzer2018ltmle} for technical derivations of the underpinnings of TMLE in the present context. 

From a high level, our estimator is built in several stages. In the first stage (Section \ref{est_prop_scores}), we obtain estimates of randomization probabilities and the probability of a participant attending each clinic visit. Collectively, we refer to these probabilities as the \emph{propensity scores}. These propensity scores, in particular, the components pertaining to missingness probabilities are used in the procedure to control for the possibility of time-varying confounding based on measured participant characteristics. 

With estimates of the propensity scores in hand, we proceed to estimating the outcome process using a sequence of regressions and starting at the final timepoint. Each of these outcome regressions involves two steps: obtaining an initial regression estimate and augmenting this estimate using fluctuation submodels. The purpose of the first step is to obtain an estimator that adjusts for potential time-varying confounders of the outcome and missingness process. The purpose of the second step is to ensure that the efficient influence function estimating equation is satisfied, which in turn endows the TMLE with desirable robustness and efficiency properties. Notably, the TMLE controls for potential time-varying confounding in two ways -- through estimation of propensity scores and through sequential estimation of the outcome process. The augmentation step is the crucial step in wedding these two approaches. In particular, this step ensures that, so long as either the outcome process or the propensity scores are consistently estimated, the resultant TMLE is consistent. That is, it ensures the estimator is \emph{doubly robust}. For more on double-robustness in the present context, see \cite{bangrobins:2005:biometrics,vdlgruber:2012:ijb}.

To provide a concrete example, below we consider that $W_0, W_1$, and $W_2$ are scalar-valued. 

\subsection{Estimating propensity scores} \label{est_prop_scores}
Our first task is to estimate the randomization probabilities for the text + step-to-eCoaching intervention. Specifically, we require estimates of \begin{align*}
\bar{g}_{A,0} &= \mathbb{P}(A_0 = 1) \  \mbox{and} \\
\bar{g}_{A,1}(y_0, y_1) &= \mathbb{P}(A_1 = d(1, Y_0, Y_1) \mid A_0 = 1, C_1 = 1, L_0 = \ell_0, Y_1 = y_1) \ . 
\end{align*}
While these probabilities are known by design, our estimators will utilize empirical estimates to account for chance imbalances between intervention arms. To that end, we define \begin{align*}
  \bar{g}_{A,0n} &= \frac{1}{n} \sum_{i=1}^n \ind(A_{0i} = 1)  \ \mbox{and} \ \\ 
  \bar{g}_{A,1n}(y_0, y_1) &= \left\{ \begin{array}{cc}
    \frac{\sum_{i=1}^n \ind(A_{0i} = 1, \ C_{1i} = 1, \ A_{1i} = 3)}{\sum_{i=1}^n \ind(A_{0i} = 1, \ C_{1i} = 1)} & \mbox{$y_1 \ge y_0$ and $y_1 \ne 0$} \\
    1 & \mbox{else} 
  \end{array}    \right. \ . 
\end{align*}
Note that the latter estimator accounts for our step-up criteria. If a participant's sexual risk behaviors do not improve, then they are eligible for step up and may be re-randomized to the eCoaching intervention. We estimate this re-randomization probability using an empirical proportion. If, on the other hand, a participant's sexual risk behaviors improve, then that participant is not eligible for step up and stays on the text intervention with probability 1. 

The second step is to estimate the probability of missing each clinic visit as a function of past covariates. In particular, for each covariate history $\bar{\ell}_2$, we require an estimate of \begin{align*}
\bar{g}_{C,1}(\ell_0) &= \mathbb{P}(C_1 = 1 \mid A_0 = 1, L_0 = \ell_0) \ , \\
\bar{g}_{C,2}(\bar{\ell}_1) &= \mathbb{P}(C_2 = 1 \mid A_0 = 1, C_1 = 1, A_1 = d(1, y_0, y_1), \bar{L}_1 = \bar{\ell}_1) \ , \ \mbox{and} \\
\bar{g}_{C,3}(\bar{\ell}_2) &= \mathbb{P}(C_3 = 1 \mid A_0 = 1, C_1 = 1, A_1 = d(1, y_0, y_1), C_2 = 1, \bar{L}_2 = \bar{\ell}_2) \ . 
\end{align*}
These estimates can be obtained, for example, using logistic regression. Considering estimation of $\bar{g}_{C,1}$, we could posit a logistic regression model, and regress the binary outcome $C_1$ onto basis functions of $L_0$ amongst participants randomized to the text intervention. Alternatively, we could posit a logistic regression model for the probability of attending the first clinic visit as a function of $A_0$ \emph{and} $L_0$, and regress $C_1$ onto basis functions of $L_0$ and $A_0$ amongst all participants. As a concrete example, consider the main-terms logistic regression model for $\tilde{g}_{C,1}(a_0, \ell_0) = \mathbb{P}(C_1 = 1 \mid A_0 = a_0, L_0 = \ell_0), $\[
  \mbox{logit}\{\tilde{g}_{C,1}(a_0, \ell_0)\} = \gamma_0 + \gamma_1 \ind(a_0 = 1) + \gamma_2 \ind(a_0 = 2) + \ell_0 \gamma_{L} \ , \ \gamma \in \mathbb{R}^5  \ ,
\]
where $\gamma = (\gamma_0, \gamma_1, \gamma_2, \gamma_L)^{\top}$ and $\gamma_L$ is a two-dimensional vector of parameters associated with $W_0$ and $Y_0$. We note that this regression \emph{pools} across interventions to estimate the conditional probability of attending the first clinic visit. An estimate $\hat{\gamma}$ of $\gamma$ can be obtained via maximum likelihood. We use $\tilde{g}_{C,1n}(a_0, \ell_0)$ to denote the fitted value from this regression for an observation with $A_0 = a_0$ and $L_0 = \ell_0$. For any $\ell_0$ our estimate of $\bar{g}_{C,1}$ is \[
  \bar{g}_{C,1n}(\ell_0) = \tilde{g}_{C,1n}(1, \ell_0) = \mbox{logit}^{-1}(\hat{\gamma}_0 + \hat{\gamma}_1 + \ell_0 \hat{\gamma}_L) \ . 
\]
To estimate $\bar{g}_{C,2}$ we take a similar approach. For example, we could posit a logistic regression model for the probability of attending the second clinic visit as a function of $\bar{A}_1$ and $\bar{L}_1$, and regress $C_2$ onto basis functions of $\bar{A}_1$ and $\bar{L}_1$ using data from all participants who attended the first clinic visit. As a concrete example, consider the main-terms logistic regression model for $\tilde{g}_{C,2}(a_0, a_1, \bar{\ell}_1) = \mathbb{P}(C_2 = 1 \mid A_0 = a_0, C_1 = 1, \bar{L}_1 = \bar{\ell}_1)$,  \begin{align*}
  \mbox{logit}\{\tilde{g}_{C,2}(a_0, a_1, \bar{\ell}_1)\} &= \eta_0 + \eta_1 \ind(a_0 = 1, a_1 = 1) + \eta_2 \ind(a_0 = 2, a_1 = 2) \\
  &\hspace{0.2in} + \eta_3 \ind(a_0 = 1, a_1 = 3) + \eta_4 \ind(a_0 = 2, a_1 = 3) + \bar{\ell}_1 \eta_L \ , \\ &\hspace{2.8in} \eta \in \mathbb{R}^{9} \ ,
\end{align*}
where $\eta = (\eta_0, \eta_1, \dots, \eta_4, \eta_L)$ and $\eta_L$ is a four-length vector of regression parameters associated with $Y_0, W_0, Y_1$, and $W_1$. An estimate $\hat{\eta}$ of $\eta$ can be obtained via maximum likelihood, and similarly as above, we use $\bar{g}_{C,2n}(a_0, a_1, \bar{\ell}_1)$ to denote the fitted value from this regression for an observation with $A_0 = a_0$, $A_1 = a_1$, and $\bar{L}_1 = \bar{\ell}_1$. For any $\bar{\ell}_1$ our estimate of $\bar{g}_{C,2}$ is \begin{align*}
  &\bar{g}_{C,2n}(\bar{\ell}_1) = \tilde{g}_{C,2n}(1, d(1, y_0, y_1), \bar{\ell}_1) \\
  & \hspace{0.05in} = \mbox{logit}^{-1}\left\{\hat{\eta}_0 + \hat{\eta}_1 \ind(d(1, y_0, y_1) = 1) + \hat{\eta}_3 \ind(d(1, y_0, y_1) = 3) + \bar{\ell}_1 \hat{\eta}_L \right\} \ . 
\end{align*}
Finally, we estimate the conditional probability of attending the third clinic visit similarly as with the second to obtain $\bar{g}_{C,3n}$, an estimate of $\bar{g}_{C,3}$. 

\subsection{Estimating sequential outcome regressions} \label{outcome_regs}

We now turn to obtaining an estimate of $\bar{Q}_3$, the conditional mean of the cumulative number of risky sexual behaviors amongst participants in the text + step-to-eCoaching who attended each clinic visit given their sexual risk behaviors and covariates measured through the second clinic visit. We can write (\ref{Qbar3}) as \begin{align*}
  \bar{Q}_3(\bar{\ell}_2) &= \mathbb{E}(Y \mid A_0 = 1, A_1 = d(1, y_0, y_1), \bar{C}_3 = \bar{1}, \bar{L}_2 = \bar{\ell}_2) \\
  &= y_1 + y_2 + \mathbb{E}(Y_3 \mid A_0 = 1, A_1 = d(1, y_0, y_1), \bar{C}_3 = \bar{1}, \bar{L}_2 = \bar{\ell}_2) \\
  &\equiv y_1 + y_2 + \tilde{Q}_3(1, d(1, y_0, y_1), \bar{\ell}_2) \ ,
\end{align*}
where we defined $\tilde{Q}_3(a_0, a_1, \bar{\ell}_2) = \mathbb{E}(Y_3 \mid A_0 = a_0, A_1 = a_1, \bar{C}_3 = \bar{1}, \bar{L}_2 = \bar{\ell}_2)$. 
The second equality follows from the fact that we are conditioning on $\bar{L}_2 = \bar{\ell}_2$, which includes past sexual risk behaviors $y_1$ and $y_2$. Thus, in order to estimate $\bar{Q}_3$, it suffices to estimate $\tilde{Q}_3$. For example, since $Y_3$ is a count variable, we could posit a generalized linear model with Poisson family and log link function for $\tilde{Q}_3$. As with estimation of missingness probabilities, we will use regressions that pool over intervention groups, by regressing the outcome $Y_3$ onto basis functions of $\bar{A}_2$ and $\bar{L}_2$. To provide a concrete example, we consider the following pooled, log-linear regression model for $\tilde{Q}_3$, \begin{align*}
  \mbox{log}\{\tilde{Q}_3(a_0, a_1, \bar{\ell}_2)\} &= \beta_0 + \beta_1 \ind(a_0 = 1, a_1 = 1) + \beta_2 \ind(a_0 = 2, a_1 = 2) \\
  &\hspace{0.25in} + \beta_3 \ind(a_0 = 1, a_1 = 3) + \beta_4 \ind(a_0 = 2, a_1 = 3) + \bar{\ell}_2 \beta_L \ , \\ &\hspace{3in}\beta \in \mathbb{R}^{12} \ .
\end{align*}
An estimate $\hat{\beta}$ of $\beta$ can be obtained via maximum likelihood, and we denote by $\tilde{Q}_{3n}(a_0, a_1, \bar{\ell}_2)$ the fitted value from this regression for an observation with $\bar{A}_1 = \bar{a}_1$ and $\bar{L}_2 = \bar{\ell}_2$. For any $\bar{\ell}_2$ an estimate of $\bar{Q}_{3}(\bar{\ell}_2)$ is \begin{align}
  \bar{Q}_{3n}(\bar{\ell}_{2}) &= y_1 + y_2 + \tilde{Q}_{3n}(1, d(1, y_0, y_1), \bar{\ell}_2) \label{Qbar3_def} \\ 
  &= y_1 + y_2 + \mbox{exp}\left\{\hat{\beta}_0 + \hat{\beta}_1 \ind(d(1, y_0, y_1) = 1) \right. \notag \\ 
  &\hspace{1.5in} + \hat{\beta}_3 \ind(d(1, y_0, y_1) = 3) + \bar{\ell}_2 \hat{\beta}_L \left. \right\} \ \notag . 
\end{align}

Next, we augment our initial estimate of $\tilde{Q}_3$ by estimating the parameter of a \emph{fluctuation submodel} associated with the \emph{augmentation covariate} \[
H_{3n}(a_0, a_1, \bar{c}_3, \bar{\ell}_2) = \frac{\ind(a_0 = 1, a_1 = d(1, y_0, y_1), \bar{c}_3 = \bar{1})}{\bar{g}_{A,0n} \ \bar{g}_{A,1n}(y_0, y_1) \ \prod_{t=1}^3 \bar{g}_{C,tn}(\bar{\ell}_{t-1})} \ .
\]
Recall, that TMLE is a general purpose tool for generating estimators that solve user-specified equations. In the current problem, we need an estimate $\bar{Q}_{3n}^*$ of $\bar{Q}_3$ that satisfies \begin{equation} \label{eif_t3}
  \frac{1}{n} \sum_{i=1}^n H_{3n}(A_{0i}, A_{1i}, \bar{C}_{3i}, \bar{L}_{2i}) \{Y - \bar{Q}_{3n}^*(\bar{L}_{2i})\} = 0 \ . 
\end{equation}
This can be achieved as follows. First, we compute maximum observed value of $Y_3$, denoted by $s_{3[n]}$ and we scale $Y_3$ to the unit interval, $Y^s_3 = Y_3/s_{3[n]}$. Next, we define the augmented regression model, \[
  \mbox{logit}\left\{ \tilde{Q}_{3}(a_0, a_1, \bar{\ell}_2)/s_{3[n]} \right\} = \mbox{logit}\{\tilde{Q}_{3n}^s(a_0, a_1, \bar{\ell}_2)/s_{3[n]}\} + \epsilon_3 H_{3n}(a_0, a_1, \bar{c}_3, \bar{\ell}_2) \ , \ \epsilon_3 \in \mathbb{R} \ . 
\]
Estimating the parameter $\epsilon_3$ corresponds to fitting a logistic regression in subset of participants $\bar{C}_2 = \bar{1}$, where the outcome is $Y^s_{3}$, the regressor is $H_{3n}$, and the model includes an offset equal to the logit of the scaled fitted value. Denoting by $\epsilon_{3n}$ the estimate of $\epsilon_3$, the augmented estimate of $\tilde{Q}_3$ is \[
\tilde{Q}_{3n}^*(a_0, a_1, \bar{\ell}_2) = s_{3[n]} \mbox{logit}^{-1}\left[ \mbox{logit}\{\tilde{Q}_{3n}(\bar{\ell}_2)/s_{3[n]}\} + \epsilon_{3n} H_{3n}(a_0, a_1, \bar{1}, \bar{\ell}_2)  \right] \ .
\]
Thus, for a given $\bar{\ell}_2$, the augmented estimate of $\bar{Q}_3(\bar{\ell}_2)$ is \[\bar{Q}_{3n}^*(\bar{\ell}_2) = y_1 + y_2 + \tilde{Q}_{3n}^*(1, d(1, y_0, y_1), \bar{\ell}_2) \ . \] It is straightforward to show that $\bar{Q}_{3n}^*$ satisfies equation (\ref{eif_t3}), as required. 

Our next task it to obtain an estimate of $\bar{Q}_2$, the conditional mean of $\bar{Q}_3$ amongst participants in the text + step-to-eCoaching intervention who attended clinic visits one and two, and given covariate-history measured through the first follow-up clinic visit. We can write (\ref{Qbar2}) as \begin{align*}
  \bar{Q}_2(\bar{\ell}_1) &= \mathbb{E}\{\bar{Q}_2(\bar{L}_2) \mid A_0 = 1, A_1 = d(1, y_0, y_1), \bar{C}_2 = \bar{1}, \bar{L}_1 = \bar{\ell}_1\} \\
  &= y_1 + \mathbb{E}\{Y_2 + \tilde{Q}_3(A_0, A_1, \bar{L}_2) \mid A_0 = 1, A_1 = d(1, y_0, y_1), \bar{C}_2 = \bar{1}, \bar{L}_1 = \bar{\ell}_1\} \\
  &\equiv y_1 + \tilde{Q}_2(1, d(1, y_0, y_1), \bar{\ell}_1) \ , 
\end{align*}
where we defined $\tilde{Q}_2(a_0, a_1, \bar{\ell}_1) = \mathbb{E}\{\tilde{Q}_3(A_0, A_1, \bar{L}_2) \mid A_0 = a_0, A_1 = a_1, \bar{C}_2 = \bar{1}, \bar{L}_1 = \bar{\ell}_1\}$. Thus, to estimate $\bar{Q}_2$, it suffices to estimate $\tilde{Q}_2$. This quantity could be estimated, as above, using a pooled log-linear regression model, \begin{align*}
  \mbox{log}\{\tilde{Q}_2(a_0, a_1, \bar{\ell}_1)\} &= \nu_0 + \nu_1 \ind(a_0 = 1, a_1 = 1) + \nu_2 \ind(a_0 = 2, a_1 = 2) \\
  &\hspace{0.2in}  + \nu_3 \ind(a_0 = 1, a_1 = 3) + \nu_4 \ind(a_0 = 2, a_1 = 3) + \bar{\ell}_1^{\top} \nu_L \ , \ \nu \in \mathbb{R}^{9} \ .
\end{align*}
In order to appropriately pool over the intervention groups, we estimate the parameters of this regression using a stacked data set with five rows for each original observation, for a total of $5n$ rows. We evaluate the fitted value $\tilde{Q}_{3n}^*(a_0, a_1, \bar{L}_{1i})$ and set the stacked outcome $Y_{2i} + \tilde{Q}_(a_0, a_1,\bar{L}_{1i})$ for $(a_0, a_1) \in \{(1, d(1, Y_{0i}, Y_{1i})), (2, d(2, Y_{0i}, Y_{1i})), (1, 1), (2, 2), (0, 0)\}$. Thus, the first $n$ rows of the stacked data set correspond to fitted values with interventions set to $(1, d(1, Y_0, Y_1))$, the second $n$ rows correspond to fitted values with interventions set to $(2, d(2, Y_0, Y_1))$, and so on. Accordingly, the first $n$ entries in the stacked $A_0$ column equal to 1, the next $n$ equal to 2, the next $n$ equal to 1, the next $n$ equal to 2, the final $n$ equal to 0. Similarly, the first $n$ entries in the stacked $A_1$ column equal to $d(1, Y_{0i}, Y_{1i}), i = 1,\dots,n$, the next to $d(2, Y_{0i}, Y_{1i}), i = 1,\dots,n$, and so on. All previously measured covariates, $(W_0, Y_0, W_1, Y_1)$, are copied five times to complete the stacked data set. 

An estimate $\hat{\nu}$ of $\nu$ can be obtained by fitting the log-linear regression model above to the subset of the stacked data with $\bar{C}_2 = \bar{1}$. We denote by $\tilde{Q}_{2n}(a_0, a_1, \bar{\ell}_2)$ the fitted value from this regression for an observation with $\bar{A}_1 = \bar{a}_1$ and $\bar{L}_1 = \bar{\ell}_1$. Thus, our estimate of $\bar{Q}_{2}$ is \begin{align*}
  \bar{Q}_{2n}(\bar{\ell}_{1}) &= y_1 + \tilde{Q}_{2n}(1, d(1, y_0, y_1), \bar{\ell}_1) \\ 
  &= y_1 + \mbox{exp}\left\{\hat{\nu}_0 + \hat{\nu}_1 \ind(d(1, y_0, y_1) = 1) + \hat{\nu}_3 \ind(d(1, y_0, y_1) = 3) + \bar{\ell}_1 \hat{\nu}_L \right\} \ . 
\end{align*}

As above, the next step is to augment the initial estimate of $\tilde{Q}_2$ using an appropriate fluctuation submodel. To that end, we define the augmentation covariate \[ 
H_{2n}(a_0, a_1, \bar{c}_2, \bar{\ell}_1) = \frac{\ind(a_0 = 1, a_1 = d(1, y_0, y_1), \bar{c}_2 = \bar{1})}{\bar{g}_{A,0n} \ \bar{g}_{A,1n}(y_0, y_1) \ \prod_{t=1}^2 \bar{g}_{C,tn}(\bar{\ell}_{t-1})} \ .
\]
The goal of this augmentation step is to generate an estimate $\bar{Q}_{2n}^*$ of $\bar{Q}_2$ that satisfies \begin{equation} \label{eif_t2}
  \frac{1}{n} \sum_{i=1}^n H_{2n}(A_{0i}, A_{1i}, \bar{C}_{2i}, \bar{L}_{2i}) \{\bar{Q}_{3n}^*(\bar{L}_{2i}) - \bar{Q}_{2n}^*(\bar{L}_{1i})\} = 0 \ . 
\end{equation}
This can be achieved using a similar strategy as above. First, we compute the maximum observed value of $Y_2 + \tilde{Q}_{3n}^*(1, d(1, Y_0, Y_1), \bar{L}_{2})$, say $s_{2n}$. We then generate a scaled outcome $Y^s_{2} = \{Y_2 + \tilde{Q}_{3n}(1, d(1, Y_0, Y_1), \bar{L}_2)\}/s_{2n}$. Next, we define an augmented regression model for $\tilde{Q}_2$, \[
  \mbox{logit}\left\{ \tilde{Q}_{2}(a_0, a_1, \bar{\ell}_1)/s_{2n} \right\} = \mbox{logit}\{\tilde{Q}_{2n}(a_0, a_1, \bar{\ell}_1) / s_{2n}\} + \epsilon_2 H_{2n}(a_0, a_1, \bar{c}_2, \bar{\ell}_1) \ , \ \epsilon_2 \in \mathbb{R} \ . 
\]
Estimating the parameter $\epsilon_2$ corresponds with fitting a logistic regression in the subset of participants with $\bar{C}_2 = \bar{1}$, where the regression outcome is $Y^s_{2}$, the single covariate is $H_{2n}$, and the model includes an offset equal to the logit of the scaled fitted value. Denoting by $\epsilon_{2n}$ the estimate of $\epsilon_2$, the augmented estimate of $\tilde{Q}_2$ is \[
\tilde{Q}_{2n}^*(a_0, a_1, \bar{\ell}_1) = s_{2n} \mbox{logit}^{-1}\left[ \mbox{logit}\{\tilde{Q}_{2n}(a_0, a_1, \bar{\ell}_1)/s_{2n}\} + \epsilon_{2n} H_{2n}(a_0, a_1, \bar{c}_2, \bar{\ell}_1)  \right] \ .
\]
Thus, the augmented estimate of $\bar{Q}_2(\bar{\ell}_1)$ is $\bar{Q}_{2n}^*(\bar{\ell}_1) = y_1 + \tilde{Q}_{2n}^*(1, d(1, y_0, y_1), \bar{\ell}_1)$. It is straightforward to show that $\bar{Q}_{3n}^*$ and $\bar{Q}_{2n}^*$ together satisfy equation (\ref{eif_t2}). 

Our next task it to obtain an estimate of $\bar{Q}_1$, the baseline-covariate-conditional mean of $\bar{Q}_2$ amongst participants initially in the text intervention who attended clinic visit one. We can write (\ref{Qbar1}) as \begin{align*}
  \bar{Q}_1(\ell_0) &= \mathbb{E}\{\bar{Q}_2(\bar{L}_1) \mid A_0 = 1, C_1 = 1, L_0 = \ell_0\} \\
  &= \mathbb{E}\{Y_1 + \tilde{Q}_2(1, d(1, Y_0, Y_1), \bar{L}_1) \mid A_0 = 1, C_1 = 1, L_0 = \ell_0\} \\
  &\equiv \tilde{Q}_1(1, \ell_0) \ , 
\end{align*}
where we defined $\tilde{Q}_1(a_0, \ell_0) = \mathbb{E}\{Y_1 + \tilde{Q}_2(1, d(1, Y_0, Y_1), \bar{L}_1) \mid A_0 = a_0, C_1 = 1, L_0 = \ell_0\}$. Obtaining an initial, and subsequently, an augmented estimate of $\bar{Q}_1$ proceeds similarly as with $\bar{Q}_2$. We first obtain an initial estimate of $\tilde{Q}_1$, e.g., through log-linear regression on a stacked data set. To generate this stacked data set, we evaluate the fitted value $\tilde{Q}_{2n}^*(a_0, a_1, \bar{L}_{1i})$ and define the stacked outcome $Y_{1i} + \tilde{Q}_2(a_0, a_1,\bar{L}_{1i})$ for \[(a_0, a_1) \in \{(1, d(1, Y_{0i}, Y_{1i})), (2, d(2, Y_{0i}, Y_{1i})), (1, 1), (2, 2), (0, 0)\} \ .\] Next, we make a stacked column of baseline intervention assignments, with the first $n$ observations equal to 1, the next $n$ equal to 2, the next $n$ equal to 1, the next $n$ equal to 2, the final $n$ equal to 0. All baseline covariates, $(W_0, Y_0)$, are copied five times to complete the stacked data set. Then, as above, we fit our regression on the stacked data set, regressing the stacked outcome on the stacked baseline intervention and covariates. Given $\ell_0$, our estimate of $\bar{Q}_1(\ell_0)$ is $\bar{Q}_{1n}(\ell_0) = \bar{Q}_{1n}(1, \ell_0)$. 

We augment $\bar{Q}_{1n}$ using the covariate \[ 
H_{1n}(a_0, c_1, \ell_0) = \frac{\ind(a_0 = 1, c_1 = 1)}{\bar{g}_{A,0n} \ \bar{g}_{C,1n}(\ell_{0})} \ ,
\]
with the goal generating an estimate $\bar{Q}_{1n}^*$ of $\bar{Q}_1$ that satisfies \begin{equation} \label{eif_t1}
  \frac{1}{n} \sum_{i=1}^n H_{1n}(A_{0i}, C_{1i}, L_{0i}) \{\bar{Q}_{2n}^*(\bar{L}_{1i}) - \bar{Q}_{1n}^*(\bar{L}_{0i})\} = 0 \ . 
\end{equation}
We again use logistic regression to this end. The scaled outcome is $Y_1^s = \{Y_1 + \tilde{Q}_{2n}(1, d(1, Y_0, Y_1), \bar{L}_2)\}/s_{1n}$, where $s_{1n}$ is the maximum observed value of $Y_1 + \tilde{Q}_{2n}(1, d(1, Y_{0i}, Y_{1i}), \bar{L}_{1i})$. As in previous steps, we regress the scaled outcome $Y^s_1$ onto the covariate $H_{1n}$ with offset equal to $\mbox{logit}(\bar{Q}_{1n}/s_{1n})$. 

Denoting by $\epsilon_{1n}$ the estimate of $\epsilon_1$, the augmented estimate of $\bar{Q}_1(\ell_0)$ is \[
\bar{Q}_{1n}^*(\ell_0) = s_{1n} \mbox{logit}^{-1}\left[ \mbox{logit}\{\bar{Q}_{1n}(\ell_0)/s_{1n}\} + \epsilon_{1n} H_{1n}(1, 1, \ell_0) \right] \ .
\]
It is straightforward to show that $\bar{Q}_{2n}^*$ and $\bar{Q}_{1n}^*$ together satisfy equation (\ref{eif_t2}). 

The final step is to plug-in the estimate $\bar{Q}_{1n}^*$ into right-hand-side of equation (\ref{Qbar0}). To evaluate this plug in estimator, we also need an estimate of $Q_0$, the cumulative distribution of $L_0$. For this, we use the empirical distribution $Q_n(\ell_0) = \frac{1}{n} \sum_{i=1}^n \ind(L_{0i} \le \ell_0)$. The final TMLE estimator is thus \[ 
  \psi_n^* = \int \bar{Q}_{1n}^*(\ell_0) dQ_n(\ell_0) = \frac{1}{n} \sum_{i=1}^n \bar{Q}_{1n}^*(L_{0i})  \ . 
\]

It is straightforward to obtain closed-form, Wald-style confidence intervals and hypothesis tests based on TMLE-estimated treatment effects using the calculus of influence functions and the functional delta method \cite{van2000asymptotic}. We provide more information on these methods in the Appendix. 

\subsection{Note on propensity score and outcome regression estimators}

In the above description of TMLE, we used initial estimators based on parametric models to illustrate key ideas. However, TMLE methodology is in no way limited to such estimators. In particular, we may wish to consider more flexible regression techniques in order to mitigate the risk of residual bias incurred by regression model misspecification. To that end, it may be propitious to utilize super learning \cite{van2007super}, a generalization of regression stacking \cite{wolpert1992stacked,breiman1996stacked}. A super learner can build estimators of regression quantities by considering all possible convex combinations of a library of pre-specified regression estimators. That is to say, a fitted value from a super learner regression is a weighted combination of fitted values from various candidate regression estimators. The candidate regression estimators can range from parametric regression estimators to highly data-adaptive estimators. The super learner selects weights assigned to each candidate estimator by finding the convex weights that minimize a cross-validated risk criteria, for example, cross-validated mean squared-error. The use of super learner is particularly appealing in the context of randomized trials in that the method can be fully pre-specified and is essentially a nonparametric method so long as sufficiently flexible regressions are included in the library. For these reasons, in the analysis of the TechStep trial, we will use the super learner to estimate the conditional probabilities of missingness and the sequential outcome regressions described in the previous paragraph. 

\section{Simulation}
We performed extensive simulations to determine the sample size required to detect clinically meaningful differences in average risk behaviors in the TechStep trial. For simplicity, we assumed $W_0$ was scalar-valued and drawn from a Bernoulli(1/2) distribution. Baseline sexual risk behaviors $Y_0$ were drawn from a Poisson(exp($\gamma_0$)) distribution, where $\gamma_0$ is the log-average number of risk behaviors. Using estimates from published data \cite{kann2017youth,reisner2019situated}, we set $\gamma_0 = \mbox{log}(1.5)$. The treatment $A_0$ was assigned randomly with 1/3 probability of control, text-, and webapp-based interventions. Data for the 3, 6, and 9-month visits were generated as follows. For $t = 1, 2, 3$, we drew $C_{t}$ from Bernoulli(logit$^{-1}(\alpha_0)$) distribution; thus, $\alpha_0$ controls the level of participant dropout. Given $C_{t} = 1$ (i.e., that the participant attended clinic visit $t$), we drew $Y_t$ from a Poisson distribution with conditional mean $(Y_{t-1} \vee \frac{1}{5}) \mbox{exp}\{\gamma_{A,1} \ind(A_{(t-1) \wedge 1} = 1) + \gamma_{A, 2} \ind(A_{(t-1) \wedge 1} = 2) + \gamma_{A, 3} \ind(A_{(t-1) \wedge 1} = 3) + \gamma_W W_0 \}$. Here, $x_1 \vee x_2$ and $x_1 \wedge x_2$ denote the smaller and larger of two real-valued numbers $x_1, x_2$, respectively. Note that this corresponds to a Markov model, where the only factors of participants' history that influence their risk behaviors are their most recently measured risk behaviors ($Y_{t-1}$), their most recent treatment ($A_{(t-1) \wedge 1}$), and baseline covariate $W_0$. We consider a wide range of scenarios by selecting different values of simulation parameters (Table \ref{sim_params}). For each combination of simulation parameters, we analyzed five hundred data sets. Power for each choice of $n$ and for each of the comparisons shown in Table \ref{causal_estimands} was estimated by the proportion of simulations in which the two-sided level 0.05 Wald test rejected the null hypothesis of no difference in average number of risky behaviors. To estimate the conditional probability of censoring, we used main terms logistic regression. To estimate the sequential outcome regression, we used the super learner based on five-fold cross-validation with a library of candidate regressions that consisted of an intercept-only model, a Poisson regression model, a negative binomial regression model, and the highly adaptive LASSO \cite{benkeser2016highly}. The latter is a flexible semiparametric regression estimator. 

\begin{table}
\begin{tabular}{lll}
\hline
Parameter & Values & Notes \\
\hline
$n$ & 200, 250, 300 & Sample size \\
$\gamma_1$ & 0, -0.11, -0.22, -0.36 & Text eff. on risky behaviors \\
$\gamma_2$ & 0, -0.22, -0.51 & Webapp eff. on risky behaviors \\
\multirow{2}{*}{$\gamma_3$} & 0, -0.11, -0.22, -0.36,  & \multirow{2}{*}{ECoaching eff. on risky behaviors} \\
  & -0.69$^*$, -0.92$^*$, -1.20$^*$, -1.61$^*$ & \\
$\alpha_0$ & -4.06, -3.35 & Loss-to-followup of 5\% and 10\% \\
\hline
\end{tabular}
\caption{Simulation parameter values and descriptions. Values of $\gamma_i$ are log rate ratios. $^*$ denotes that simulations were performed only when $\gamma_1 = 0, -0.11$.} 
\label{sim_params}
\end{table}

\begin{figure}
\includegraphics[width=\textwidth]{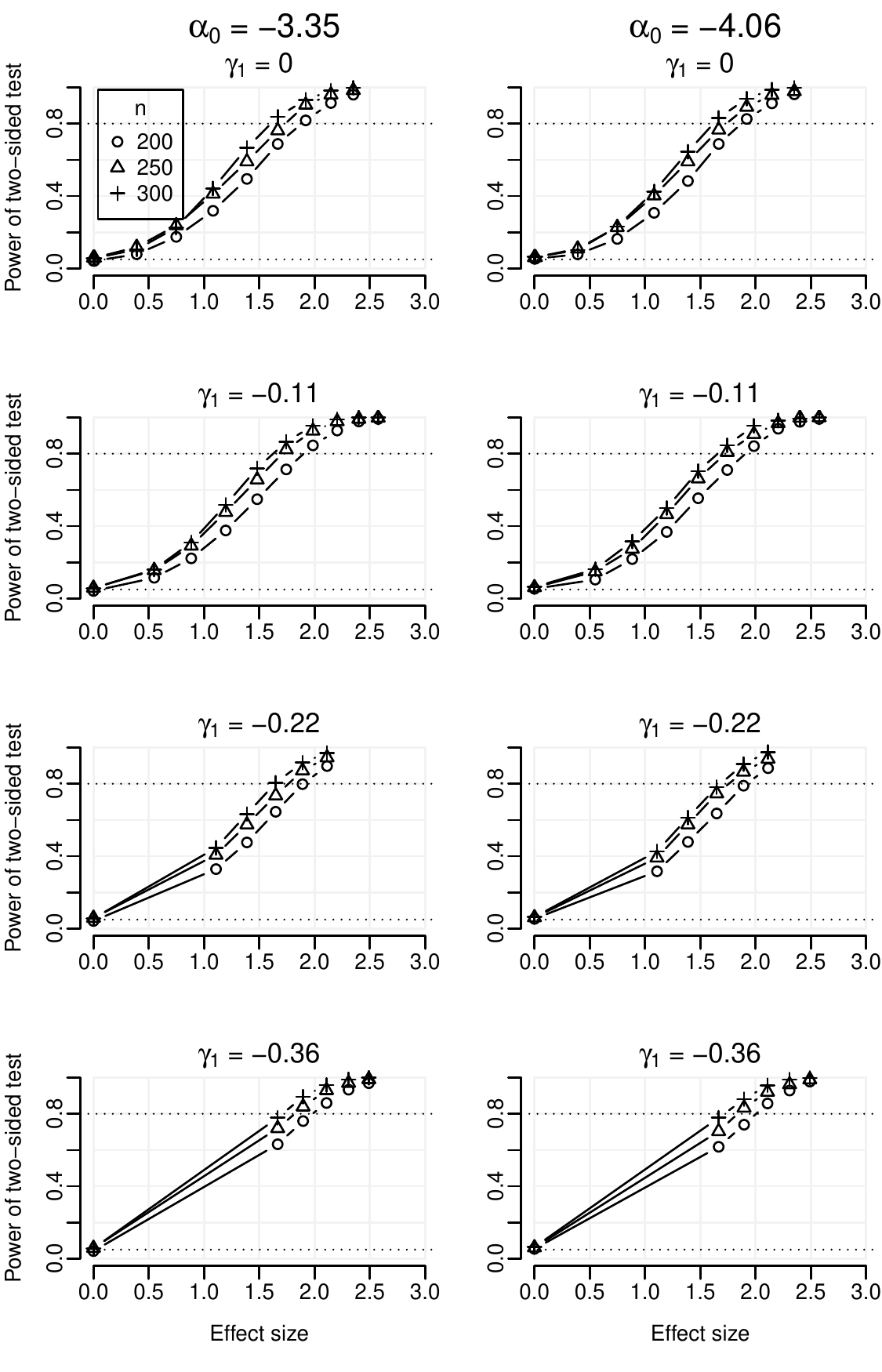}
\caption{Probability of rejecting $H_0: \mathbb{E}[Y^{\text{(ii)}}] = \mathbb{E}[Y^{\text{(i)}}]$ (no effect of text-step versus control) as a function of the size of the effect, $\mathbb{E}[Y^{\text{(ii)}}] - \mathbb{E}[Y^{\text{(i)}}]$. Dotted lines indicate the nominal type I error rate of the Wald test (0.05) and power of 80\%.}
\label{fig:sim1}
\end{figure}

\begin{figure}
\includegraphics[width=\textwidth]{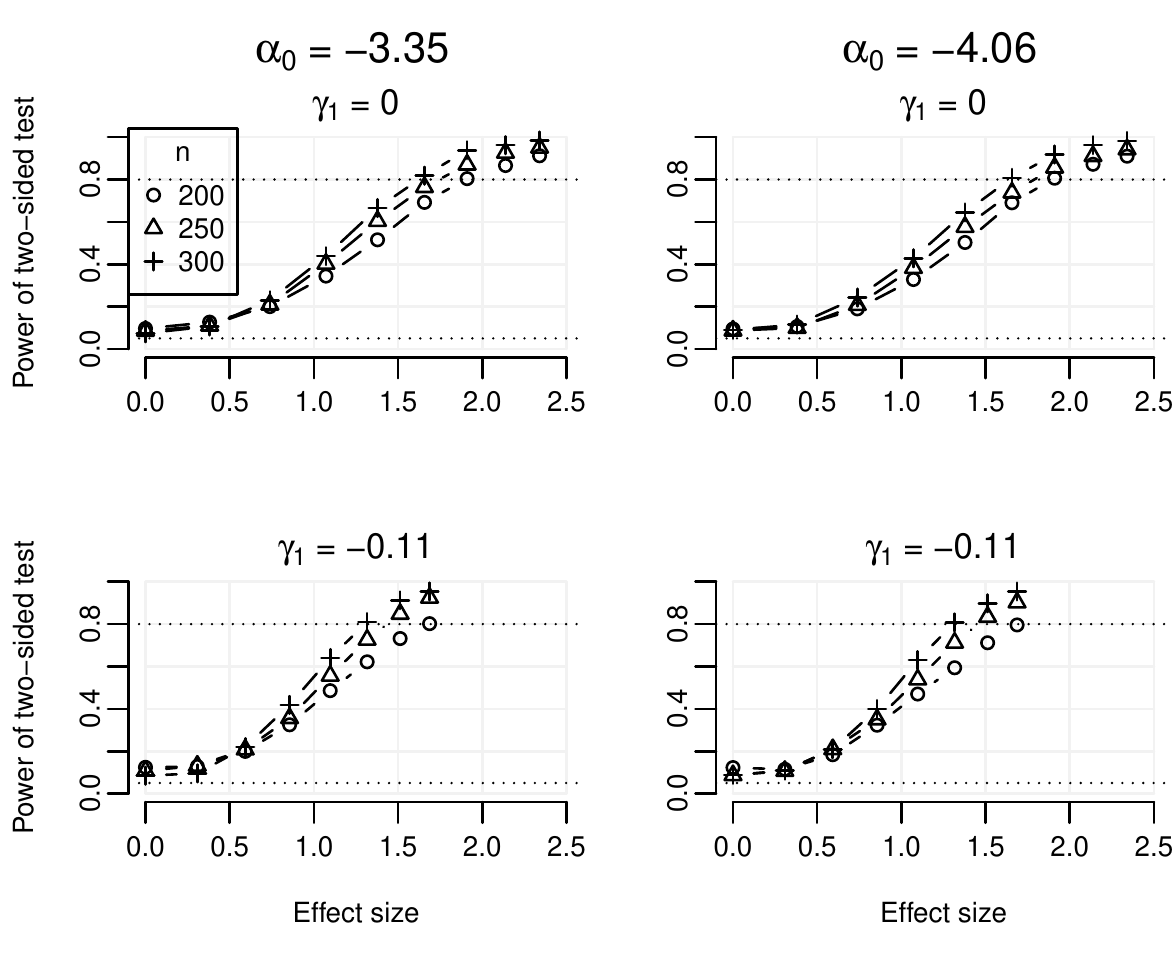}
\caption{Probability of rejecting $H_0: \mathbb{E}[Y^{\text{(ii)}}] = \mathbb{E}[Y^{\text{(iia)}}]$ (no effect of text + step versus text-only) as a function of the size of the effect, $\mathbb{E}[Y^{\text{(ii)}}] - \mathbb{E}[Y^{\text{(iia)}}]$. Dotted lines indicate the nominal type I error rate of the Wald test (0.05) and power of 80\%.}\label{fig:sim2}
\end{figure}

\begin{figure}
\includegraphics[width=\textwidth]{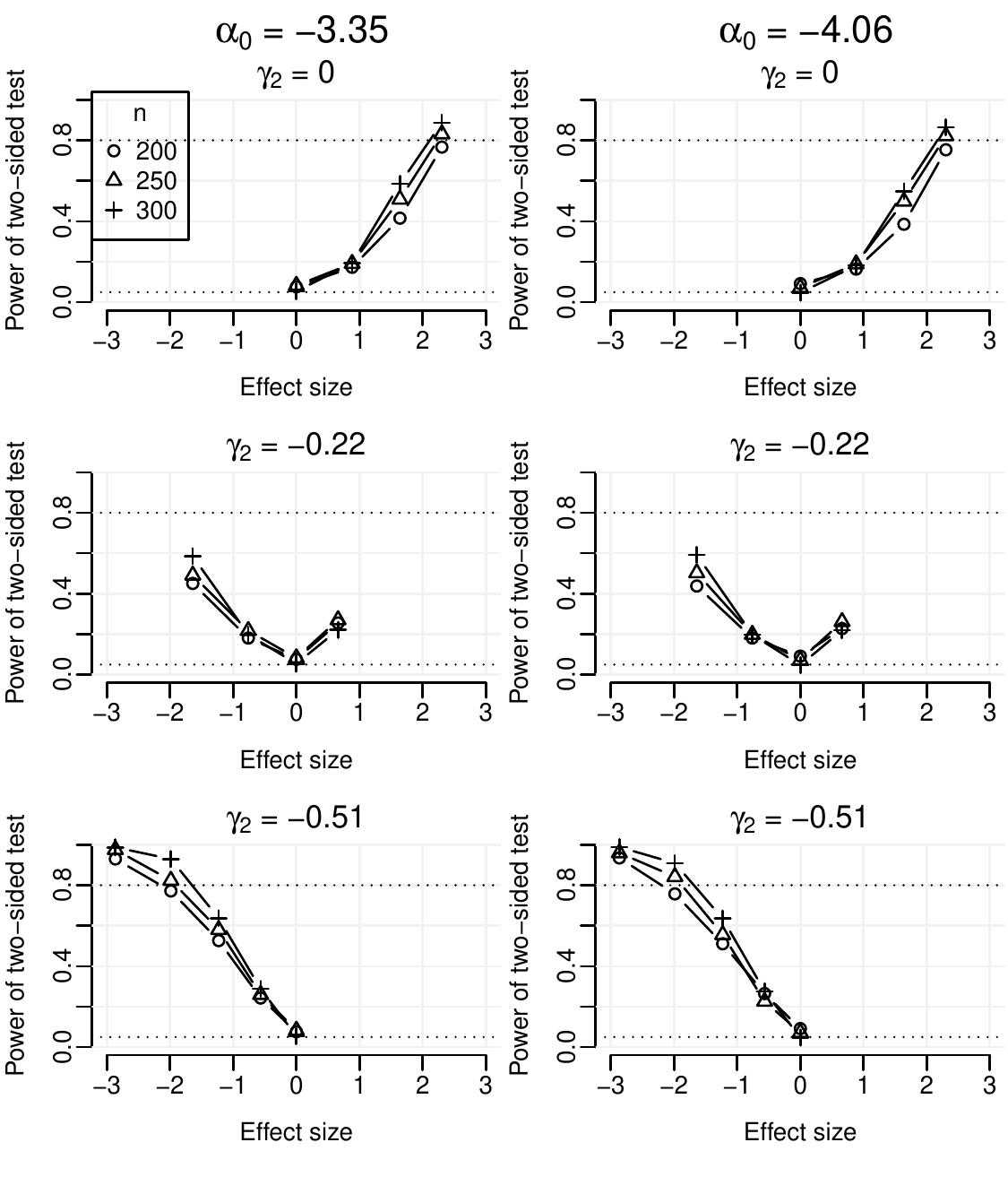}
\caption{Probability of rejecting $H_0: \mathbb{E}[Y^{\text{(ii)}}] = \mathbb{E}[Y^{\text{(iii)}}]$ (no effect of text-only versus app-only) as a function of the size of the effect, $\mathbb{E}[Y^{\text{(ii)}}] - \mathbb{E}[Y^{\text{(iii)}}]$. Dotted lines indicate the nominal type I error rate of the Wald test (0.05) and power of 80\%.}\label{fig:sim3}
\end{figure}

For the primary analysis, we found that type I error rate was adequately controlled for each sample size considered (Figure \ref{fig:sim1}). Across a wide range of settings, we estimated that we have 80\% power to detect a difference in 1.60 average risk behaviors comparing text+step versus control at $n = 300$, 1.75 at $n = 250$, and 1.90 at $n = 200$. These results seem to be robust across situations where the majority of the benefit is derived from the text versus from the eCoaching component of the intervention. Increasing censoring from 5\% to 10\% (left versus right column of Figure \ref{fig:sim1}) tended to decrease power by 3-5\%. For the secondary analysis examining the effect of text+step versus text-only interventions we found that type I error rates were slightly higher than the nominal level in the smallest sample size (Figure \ref{fig:sim2}). We found similar power curves for this secondary analysis as with the primary analysis. There tended to be higher power to detect differences when there was a small effect of the text intervention ($\gamma_1 = -0.11$) relative to no effect of the text intervention ($\gamma_1 = 0$). Results for the simulation studying the secondary analysis that compares efficacy of text-only versus app-only are included in the supplementary material. These results were similar to the primary analysis with 80\% power to detect differences in average number of risk behaviors of between 1.60 and 1.90. Results for comparing text-only versus app-only interventions are shown in Figure \ref{fig:sim3}. We found a minor increase in type I error for this comparison using the proposed test. Across the various settings, there is 80\% power to detect an average difference of about two risk behaviors between the text-only and app-only interventions. 

\section{Discussion}

Based on these simulation results and other criteria, a target of $n=250$ was set for enrollment in the TechStep trial. Our simulation results suggest this sample size will afford sufficient power to detect clinically meaningful differences between interventions, and balances power appropriately between the primary analyses and key secondary analyses. 

Further research is warranted in several areas. First, our Wald tests were slightly anti-conservative for testing differences between the text+step versus text condition. We will explore alternative standard error estimators including cross-validated influence function-based standard errors. Second, we may wish to evaluate whether and how to adjust for intervention engagement (e.g., webapp paradata) in the primary and secondary analyses. If such information is predictive of risky behaviors, then incorporating such information may lead to improved power to detect differences amongst the interventions. Finally, we will explore different techniques for time-varying confounder selection. Extensive surveys will be collected at each clinic visit, so $W_0, W_1, W_2$ are all potentially high-dimensional vectors. We will evaluate different strategies for selecting components to include in the primary and secondary analyses. 

\section*{Appendix}

\subsection*{Confidence intervals and hypothesis tests based on influence functions}

We note that our specific labeling of interventions (e.g., $A_0 = 1$ correpsonds to text, $A_0 = 2$ to app, etc...) is arbitrary. Moreover, by redefining the rule $d(a_0, y_0, y_1)$ (e.g., to equal a constant for any value of $a_0$), then we can directly apply the methodology described in the main text to estimate the counterfactual outcome under any of the interventions of interest. Let $\psi^{(\cdot)}_n$ denote the TMLE of $\mathbb{E}[Y^{(\cdot)}]$. Similarly, let $\bar{Q}_{tn}^{(\cdot)}$ denote the augmented sequential outcome regression estimate and $H_{tn}^{(\cdot)}$ the augmentation covariate. Define the \emph{estimated influence function} of $\psi_n^{(\cdot)}$ evaluated on observation $i$ as $\sum_{t=0}^3 D_{ti}^{(\cdot)}$, where \begin{align*}
D_{3i}^{(\cdot)} &= H_{3n}^{(\cdot)}(A_{0i}, A_{1i}, \bar{C}_{3i}, \bar{L}_{2i}) \{Y - \bar{Q}_{3n}^{(\cdot)}(\bar{L}_{2i})\} \\
D_{2i}^{(\cdot)} &= H_{2n}^{(\cdot)}(A_{0i}, A_{1i}, \bar{C}_{2i}, \bar{L}_{1i}) \{\bar{Q}_{3n}^{(\cdot)}(\bar{L}_{2i}) - \bar{Q}_{2n}^{(\cdot)}(\bar{L}_{1i})\} \\
D_{1i}^{(\cdot)} &= H_{1n}^{(\cdot)}(A_{0i}, C_{1i}, L_{0i}) \{\bar{Q}_{2n}^{(\cdot)}(\bar{L}_{1i}) - \bar{Q}_{1n}^{(\cdot)}(L_{0i})\} \\
D_{0i}^{(\cdot)} &= \bar{Q}_{1n}^{(\cdot)}(L_{0i}) - \int \bar{Q}_{1n}^{(\cdot)}(\ell_0) dQ_n(\ell_0) \ . 
\end{align*}
An estimate of the standard error of $n^{1/2} \psi_n^{(\cdot)}$ is \[
\sigma_n^{(\cdot)} = \left[\frac{1}{n} \sum_{i=1}^n \sum_{t=0}^3 \{D_{ti}^{(\cdot)}\}^2 \right]^{1/2}
\]
and the confidence interval $\psi^{(\cdot)}_n \pm z_{1-\alpha/2} \sigma_n^{(\cdot)} / n^{1/2}$, where $z_{1-\alpha/2}$ is the $1 - \alpha/2$ quantile of a standard Normal distribution, will have asymptotic coverage no smaller than $1-\alpha$.

Similar influence function-based standard error estimates can be used to describe the asymptotic covariance between any two TMLEs. For example, the asymptotic covariance of $n^{1/2}(\psi_n^{(\text{i})}, \psi_n^{(\text{ii})})^{\top}$ is consistently estimated by \[
  \Sigma_n^{(\text{i)(ii})} = \left[\begin{array}{ll}
    \{\sigma_n^{(\text{i})}\}^2 & \sigma_n^{(\text{i})(\text{ii})} \\
    \sigma_n^{(\text{i})(\text{ii})} & \{\sigma_n^{(\text{ii})}\}^2
  \end{array} \right] \ , \ \mbox{where} \ \sigma_n^{(\text{i})(\text{ii})} = \frac{1}{n} \sum_{i=1}^n \left\{\sum_{t=0}^3 D_{ti}^{(\text{i})} \sum_{t=0}^3 D_{ti}^{(\text{ii})} \right\} \ . 
\]
Thus, by the Delta method, an estimate of the standard error of $n^{1/2}(\psi_n^{(\text{ii})} - \psi_n^{(\text{i})})$ is \[
 \tau_n = \left[\{\sigma_n^{(\text{i})}\}^2 + \{\sigma_n^{(\text{ii})}\}^2 - 2 \sigma_n^{(\text{i})(\text{ii})} \right]^{1/2} \ . 
\]
A Wald test that rejects the null hypothesis $\mathbb{E}(Y^{(\text{ii})}) = \mathbb{E}(Y^{(\text{i})})$ whenever \[
  \bigg| \frac{n^{1/2}\{\psi_n^{(\text{ii})} - \psi_n^{(\text{i})}\}}{\tau_n} \bigg| > z_{1-\alpha/2}
\]
will have asymptotic level $\alpha$.


\bibliographystyle{spmpsci}      
\bibliography{references}   

\end{document}